\documentclass[aps,prb,twocolumn,floatfix,superscriptaddress]{revtex4-1}
\usepackage{graphicx,txfonts,bm}
\usepackage[colorlinks,citecolor=magenta,linkcolor=blue]{hyperref}

\begin{document}


\title{Unraveling the 4$f$ electronic structures of cerium monopnictides}


\author{Haiyan Lu}
\affiliation{Science and Technology on Surface Physics and Chemistry Laboratory, P.O. Box 9-35, Jiangyou 621908, China}

\author{Li Huang}
\email{lihuang.dmft@gmail.com}
\affiliation{Science and Technology on Surface Physics and Chemistry Laboratory, P.O. Box 9-35, Jiangyou 621908, China}

\date{\today}


\begin{abstract}
In order to unveil the 4$f$ electronic structures in cerium monopnictides (Ce$X$, where $X$ = N, P, As, Sb, and Bi), we employed a state-of-the-art first-principles many-body approach, namely the density functional theory in combination with the single-site dynamical mean-field theory, to make detailed calculations. We find that the 4$f$ electrons in CeN are highly itinerant and mixed-valence, showing a prominent quasi-particle peak near the Fermi level. On the contrary, they become well localized and display weak valence fluctuation in CeBi. It means that a 4$f$ itinerant-localized crossover could emerge upon changing the $X$ atom from N to Bi. Moreover, according to the low-energy behaviors of 4$f$ self-energy functions, we could conclude that the 4$f$ electrons in Ce$X$ also demonstrate interesting orbital-selective electronic correlations, which are similar to the other cerium-based heavy fermion compounds.  
\end{abstract}


\maketitle


\section{Introduction\label{sec:intro}}

The cerium-based heavy fermion materials, which exhibit a variety of fascinating and exotic properties (including topology, unconventional superconductivity, quantum criticality, mixed-valence behavior, Kondo physics, and so on), have renewed a lot of interests in recent years~\cite{RevModPhys.56.755,RevModPhys.63.239}. It is generally believed that the physical and chemical properties of cerium-based heavy fermion materials are governed by their 4$f$ electronic structures, which are very sensitive to the surrounding environment, such as external pressure, temperature, element substitution, and electromagnetic field, etc. 

For instance, Ce$_{3}$Bi$_{4}$Pt$_{3}$ (a noncentrosymmetric Kondo insulator), is such an archetypal heavy fermion compound~\cite{Lai93}. Experimentally, it has been found that a phase transition from topological Kondo insulator (TKI) to Weyl-Kondo semimetal (WKSM) could be realized via simple Pt-Pd substitution, i.e., from Ce$_{3}$Bi$_{4}$Pt$_{3}$ to Ce$_{3}$Bi$_{4}$Pd$_{3}$~\cite{PhysRevLett.118.246601}. Further theoretical calculations suggest that the underlying mechanism of the TKI-WKSM transition is the large mass difference between Pt and Pd atoms, which leads to a big discrepancy in the strength of spin-orbit coupling, and thus has an unprecedentedly drastic influence on the hybridization between 4$f$ and $p$ electrons~\cite{zhu2019}. Very recently, high pressure X-ray diffraction and electrical transport measurements for Ce$_{3}$Bi$_{4}$Pt$_{3}$ reveal that uniform compression can enhance the $f-p$ hybridization and 4$f$ electron delocalization, and finally lead to closure of the Kondo gap~\cite{pag2019}. Besides chemical doping and external pressure, it is discovered that strong magnetic field is capable of suppressing the Kondo gap as well, and inducing a Landau Fermi-liquid (metallic) state in Ce$_{3}$Bi$_{4}$Pd$_{3}$~\cite{neil2019}. These experimental facts manifest that Ce$_{3}$Bi$_{4}$Pt$_{3}$ and its substitution series Ce$_{3}$Bi$_{4}$(Pt$_{1-x}$Pd$_{x}$)$_{3}$ ($0 \leq x \leq 1$) are versatile platforms for studying the interplay between topology and electronic correlation under the influence of external conditions. In addition, Ce$T$In$_{5}$ ($T$ = Co, Rh, and Ir)~\cite{shim:1615,PhysRevLett.108.016402} and Ce$M$$_{2}$Si$_{2}$ ($M$ = Ru, Rh, Pd, and Ag)~\cite{PhysRevB.98.195102} are also classic examples for examining the intriguing properties of cerium-based heavy fermion materials.

\begin{figure}[ht]
\centering
\includegraphics[width=\columnwidth]{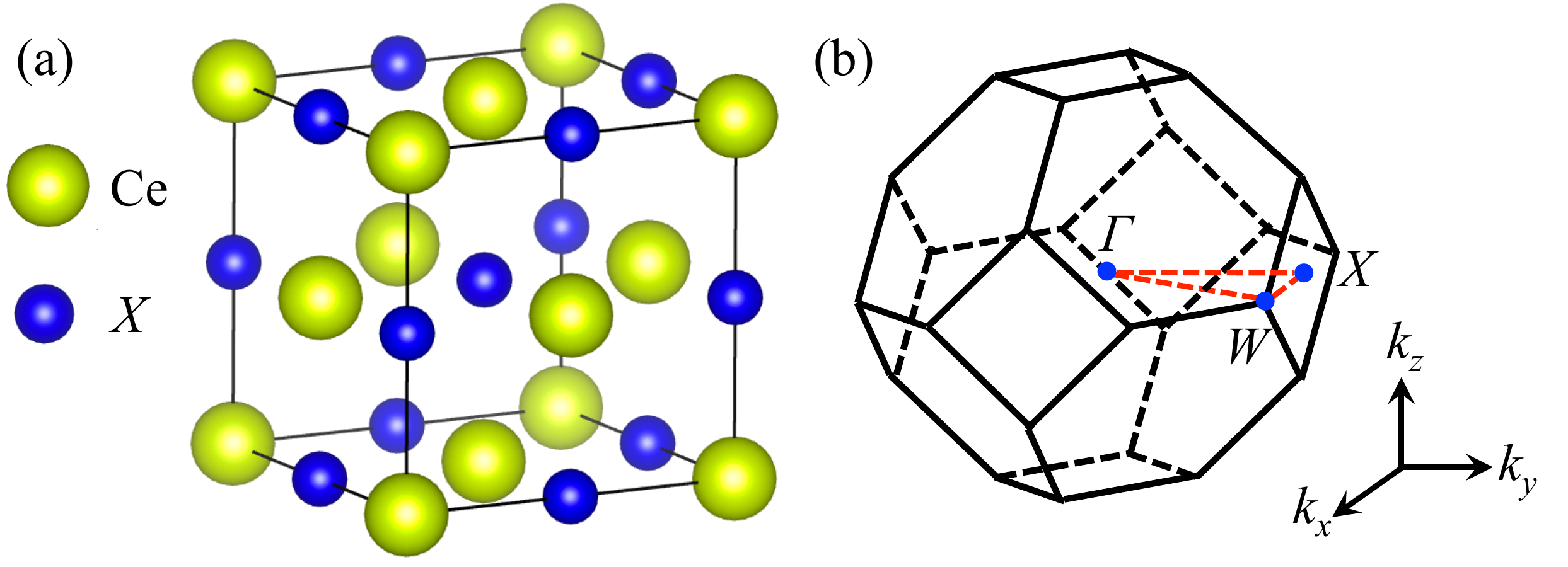}
\caption{(Color online). (a) Crystal structure of Ce$X$, where $X$ = N, P, As, Sb, and Bi. (b) Schematic picture of the first Brillouin zone of Ce$X$. Some high-symmetry $k$ points are marked.
\label{fig:tstruct}}
\end{figure}

\begin{table}[ht]
\caption{Lattice constants and key physical properties of Ce$X$~\cite{Duan2007Electronic}, where SM denotes semimetal, SC semiconductor, PM paramagnetism, AFM antiferromagnetic state, and $T_N$ the N\'{e}el temperature. Notice that Ce$X$ usually exhibits some kinds of magnetic ordering, except for CeN.
\label{tab:param}}
\begin{ruledtabular}
\begin{tabular}{ccccc}
$X$ & $a$ ({\AA}) & Metallicity & Ordering & $T_{\text{N}}$ (K) \\
\hline 
N & 5.013 & SM & PM & -\\
P & 5.942 & SC & AFM & 6$\sim$9 \\
As & 6.060 & SC & AFM & 5$\sim$7.5 \\ 
Sb & 6.400 & SM & AFM (complicated) & 16 \\
Bi & 6.490 & Metal & AFM (complicated) & 25 \\
\end{tabular}
\end{ruledtabular}
\end{table}

In the present work, let's turn to another interesting series of cerium-based heavy fermion compounds, namely cerium monopnictides Ce$X$, where $X$=N, P, As, Sb, and Bi. These compounds crystallize in the rock-salt structure (see Fig.~\ref{fig:tstruct}), in which the Ce atoms form a face-centered-cubic Bravais lattice, while the $X$ atoms occupy the octahedral voids in the lattice~\cite{Duan2007Electronic}. Owing to the peculiar electronic and magnetic properties (see Table~\ref{tab:param}), the cerium monopnictides have attracted a lot of attentions. All of the Ce$X$ compounds develop some kinds of antiferromagnetic ordering at low temperature, except for CeN. Especially, CeSb and CeBi exhibit extremely complicated magnetic ordering with large magnetic anisotropy along the [001] axis~\cite{Kohgi2000Physics}. More interestingly, CeSb even undergoes further six (magnetic or structural) phase transitions between 8~K $\sim$ 16~K~\cite{Hulliger1975Low,Fischer1978Magnetic,Meier1978Magnetic}. The exact magnetic ordering and underlying mechanism of CeSb are yet under intense debate. These low temperature magnetic phases are tightly associated with the complex electronic structures. Under cubic crystal field, 4$f_{5/2}$ state of Ce atom splits into doublet $\Gamma_7$ and quartet $\Gamma_8$ states. For CeP and CeAs, the energy gaps between the $\Gamma_7$ and $\Gamma_8$ states are about 140~K. But for CeSb and CeBi, the energy gaps are around 19~K$\sim$ 26~K~\cite{Busch1971} and 4~K$\sim$ 8~K~\cite{Birgeneau1973}, respectively. The comparatively small crystal field splitting enables the random distribution of spins for CeSb and CeBi, which could be easily influenced by temperature or magnetic field. Such anisotropic exchange interaction is regarded as an important factor for driving these magnetic transitions~\cite{Duan2007Electronic}. 

Considerable experimental progresses have been achieved on Ce$X$ compounds to disclose their electronic structures and magnetic properties. Their band structures, density of states, and Fermi surfaces have been extensively studied by using the photoemission spectroscopy (PES), angle-resolved photoemission spectroscopy (ARPES), optical conductivity, and de Haas-van Alphen (dHvA) quantum oscillation~\cite{PhysRevB.96.041120,PhysRevB.55.R3355,Kumigashira1999High,Takahashi2001High}. As for CeSb, high resolution ARPES data demonstrate the change of Fermi surface topology during the magnetic phase transition~\cite{Aoki1985Fermi,Aoki1985CeSb,PhysRevLett.78.1948,PhysRevB.56.13654,Takahashi199865,Ito2004Para,Takayama2009} and imply the dual nature of 4$f$ electrons (being itinerant or localized). On the theoretical side, the experimental photoemission spectra of Ce$X$ have been roughly reproduced by first-principles calculations~\cite{PhysRevLett.53.1673,PhysRevB.86.115116,PhysRevB.31.6251,JPSJ.76.044707,Sakai2007,Sakai2005}. Particularly, the mixed-valence nature, lattice dynamics, and elastic properties of CeN~\cite{PhysRevB.55.R10173,PhysRevB.75.045114,PhysRevB.84.205135}, the band structure and Fermi surface topology of CeSb~\cite{PhysRevB.96.035134} are well studied. In most cases, the 4$f$ electrons are assumed to be localized for heavier Ce$X$. The $f-p$ mixing model~\cite{Toshiyoshi1981,TAKEGAHARA1981857,JPSJ.72.2071} based on the anisotropic hybridization between the Ce-4$f$ level and the ligand $X$-$p$ band is widely utilized to explain the complex antiferromagnetic ordering phases~\cite{Duan2007Electronic}. However, since the 4$f$ electrons are usually correlated, the traditional first-principles approaches often underestimate the electron correlation and can not formulate a reliable physical picture of the 4$f$ electronic structures of Ce$X$. Furthermore, large spin-orbital coupling and intricate magnetic ordering states make the theoretical calculations quite difficult. Consequently, it seems tough to acquire an accurate and comprehensive description for the electronic structures of Ce$X$.

Though much effort has been devoted to understanding the unusual properties of Ce$X$ in past decades, there are still lots of issues and questions that need to be solved and answered. First of all, how do the 4$f$ electronic states evolve when $X$ goes from N to Bi? In general, the lattice constants and strength of spin-orbital coupling should vary with respect to $X$'s atomic mass. The hybridization between Ce-4$f$ and $X$-$p$ bands should be modified as well. We suspect that these changes could probably drive a 4$f$ itinerant-localized crossover or transition in this series. Second, how to explain the complicated magnetic ordering states in CeSb and CeBi? Are they related to the increment of 4$f$ electronic localization or anything else? Third, it is suggested that valence state fluctuation and orbital-dependent electronic correlation are universal features in cerium-based heavy fermion materials. We wonder whether Ce$X$ could evince similar behaviors or not. Notice that CeN was recognized as an intermediate mixed-valence compound~\cite{PhysRevLett.39.956,PhysRevB.18.4433,PhysRevB.24.3651,PhysRevB.42.8864,WACHTER2013235,Wuilloud1985}. In addition, CeP undergoes an isostructural transition ($\sim$ 8\% volume collapse) together with considerable change of 4$f$ valence state under moderate pressure~\cite{PhysRevLett.36.366}. However, we know a little about the other cerium monopnictides. In order to tackle these problems, we try to study the electronic structures of Ce$X$ thoroughly via the density functional theory merged with the single-site dynamical mean-field theory~\cite{RevModPhys.68.13}. According to the calculated results, we find that Ce$X$ is a good testing bed not only for exploring evolution of 4$f$ electronic states tuned by spin-orbital coupling, but also for studying subtle entanglement between electronic correlation and magnetism.

The rest of this paper is organized as follows. In Sec.~\ref{sec:method}, the computational details are introduced. In Sec.~\ref{sec:results}, the electronic band structures, total and partial 4$f$ density of states, hybridization functions, 4$f$ self-energy functions, and histograms of atomic eigenstates are presented. The excellent consistency between calculated and experimental data is illustrated. In Sec.~\ref{sec:dis}, we attempt to clarify some important topics about the 4$f$ itinerant-localized crossover and the possible relationship between electronic correlation and magnetic ordering states. Finally, Sec.~\ref{sec:summary} serves as a brief conclusion.  


\section{Methods\label{sec:method}}

As mentioned above, since the 4$f$ electrons in Ce$X$ are correlated, we have to consider the correlation effect carefully in the calculations. In the present work, we employed the density functional theory plus single-site dynamical mean-field theory (DFT + DMFT) approach~\cite{RevModPhys.68.13}. It incorporates the band picture inheriting from the DFT part, and a non-perturbative treatment to the 4$f$ electronic correlation from the DMFT perspective. It has been widely used to study the electronic structures of many cerium-based heavy fermion materials~\cite{Goremychkin186,PhysRevLett.112.106407,shim:1615,PhysRevLett.108.016402,PhysRevB.98.195102,PhysRevB.94.075132,PhysRevB.95.155140}. Note that the DFT + DMFT method has been applied to study Ce$X$'s electronic structures a few years ago~\cite{JPSJ.76.044707,Sakai2007,Sakai2005}. Those works using the non-crossing approximation as quantum impurity solver could reproduce the Kondo peak around the Fermi level. However, the other works employing the spin-polarized $T$-matrix fluctuation-exchange approximation and Hubbard-I approximation as quantum impurity solvers failed to capture the experimental 4$f$ states at the Fermi level of Ce$X$~\cite{PhysRevB.86.115116}.  

Here we used the \texttt{WIEN2K} code to perform the DFT calculations, which implements a full-potential augmented plane wave formalism. The experimental crystal structures of Ce$X$ are used. The muffin-tin radii for Ce and $X$ atoms are 2.5~au and 1.9~au, respectively. The $k$-points mesh for Brillouin zone integration is $21 \times 21\times 21$. The generalized gradient approximation, namely the Perdew-Burke-Ernzerhof functional~\cite{PhysRevLett.77.3865} is adopted to express the exchange-correlation potential. The spin-orbital coupling is explicitly included as well.

The basic idea of the DMFT is to map the quantum lattice model to a quantum impurity model self-consistently and solve the obtained quantum impurity model by using various quantum impurity solvers~\cite{RevModPhys.68.13}. We employ the \texttt{EDMFTF} software package~\cite{PhysRevB.81.195107} to accomplish this job. The constructed multi-orbital quantum impurity models are solved using the hybridization expansion continuous-time quantum Monte Carlo impurity solver (dubbed as CT-HYB)~\cite{RevModPhys.83.349,PhysRevLett.97.076405,PhysRevB.75.155113}. As mentioned before, the Ce-4$f$ orbitals are treated as correlated. The Coulomb repulsion interaction $U$ and the Hund's exchange interaction $J_{\text{H}}$ are 6.0 eV and 0.7 eV, respectively~\cite{PhysRevB.98.195102,PhysRevB.95.155140}. The fully localized limit (FLL) scheme~\cite{jpcm:1997} is used to describe the double counting term in 4$f$ self-energy functions. In order to simplify the calculations, we not only utilize the good quantum numbers $N$ and $J$ to reduce the sizes of matrix blocks of the local Hamiltonian, but also make a truncation for the local Hilbert space~\cite{PhysRevB.75.155113}. Only those atomic eigenstates with $N \in$ [0,3] are retained in the calculations. The lazy trace evaluation trick is used to accelerate the Monte Carlo sampling further. Since the inverse temperature $\beta = 100$ ($T \sim 116.0$~K), it is reasonable to retain only the paramagnetic solutions. We perform charge fully self-consistent DFT + DMFT calculations. Of the order of 80 DFT + DMFT iterations are required to obtain good convergence for the chemical potential $\mu$, charge density $\rho$, and total energy $E_{\text{DFT + DMFT}}$. The convergence criteria for charge and energy are $10^{-5}$ e and $10^{-5}$ Ry, respectively. The Matsubara self-energy functions $\Sigma(i\omega_n)$ generated in the last 10 DFT + DMFT iterations are collected and then averaged for further post-processing.


\section{Results\label{sec:results}}

\begin{figure*}[th]
\centering
\includegraphics[width=\textwidth]{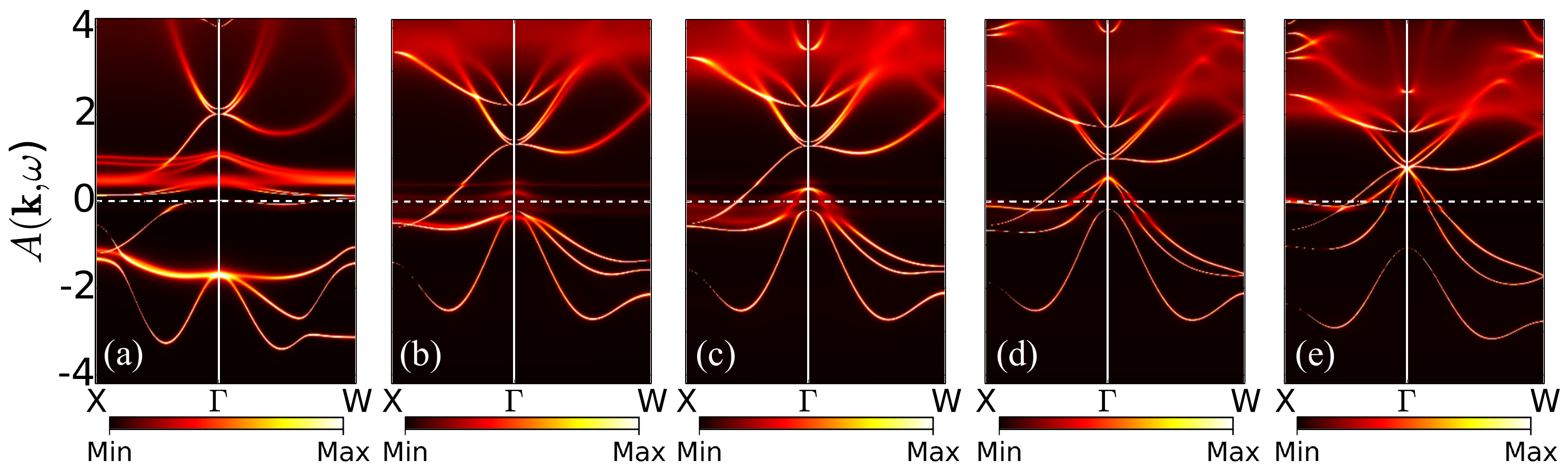}
\caption{(Color online). Momentum-resolved spectral functions $A(\mathbf{k},\omega)$ of Ce$X$ under ambient pressure and at $T =$ 116~K. The horizontal lines denote the Fermi level. (a) CeN. (b) CeP. (c) CeAs. (d) CeSb. (e) CeBi. 
\label{fig:akw}}
\end{figure*}

\begin{figure*}[th]
\centering
\includegraphics[width=\textwidth]{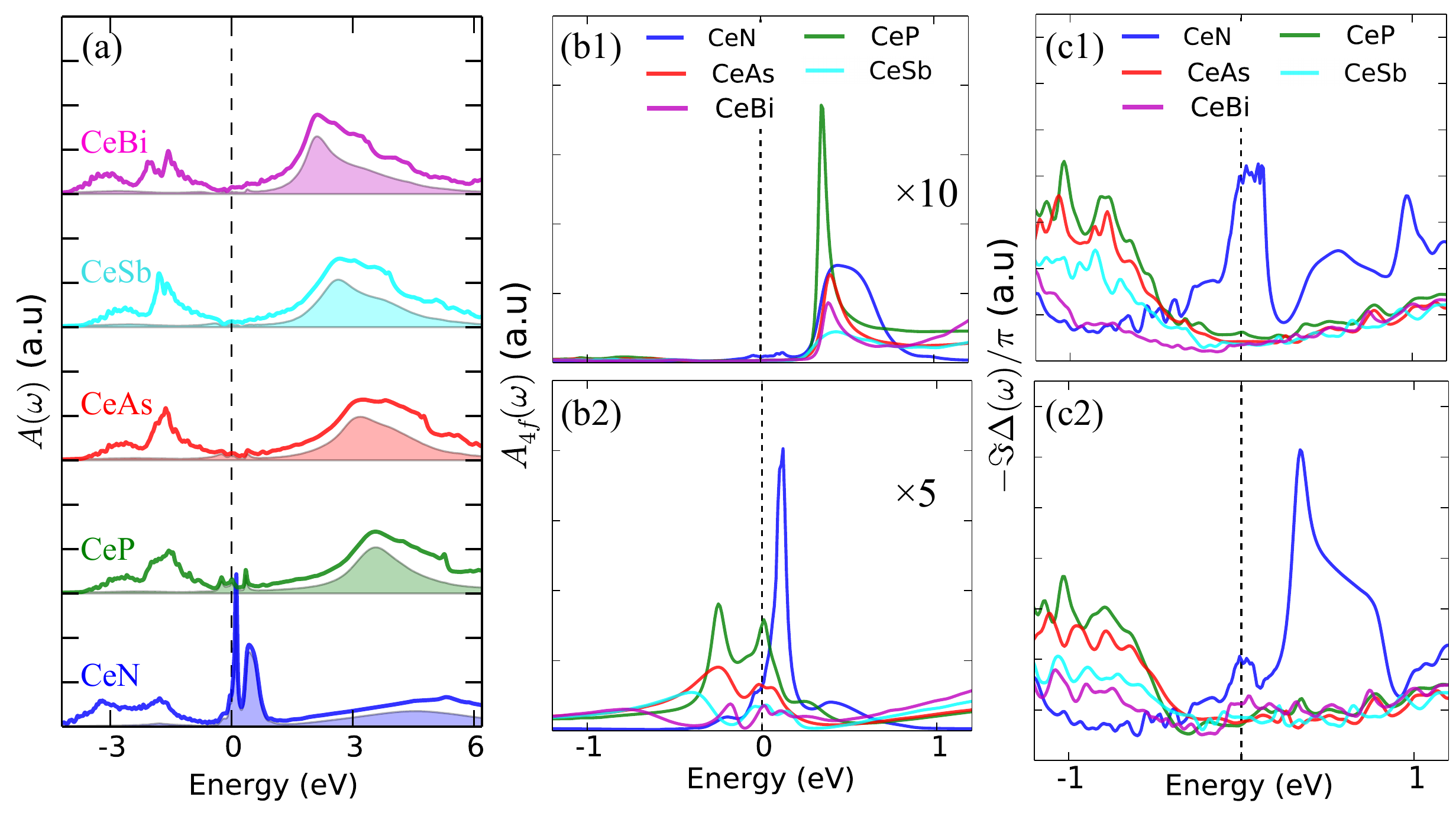}
\caption{(Color online). Electronic density of states of Ce$X$ ($X=$ N, P, As, Sb, and Bi). (a) Total density of states (thick solid lines) and partial 4$f$ density of states (color-filled regions). (b) The $j$-resolved 4$f$ partial density of states. The $4f_{5/2}$ and $4f_{7/2}$ components are represented in (b2) and (b1), respectively. (c) Imaginary parts of hybridization functions. The $4f_{5/2}$ and $4f_{7/2}$ components are depicted in (c2) and (c1), respectively. The vertical dashed lines denote the Fermi level. All of the data presented in panels (b1) and (b2) are rescaled for a better view. \label{fig:tdos}}
\end{figure*}

\begin{figure}[th]
\centering
\includegraphics[width=0.5\textwidth]{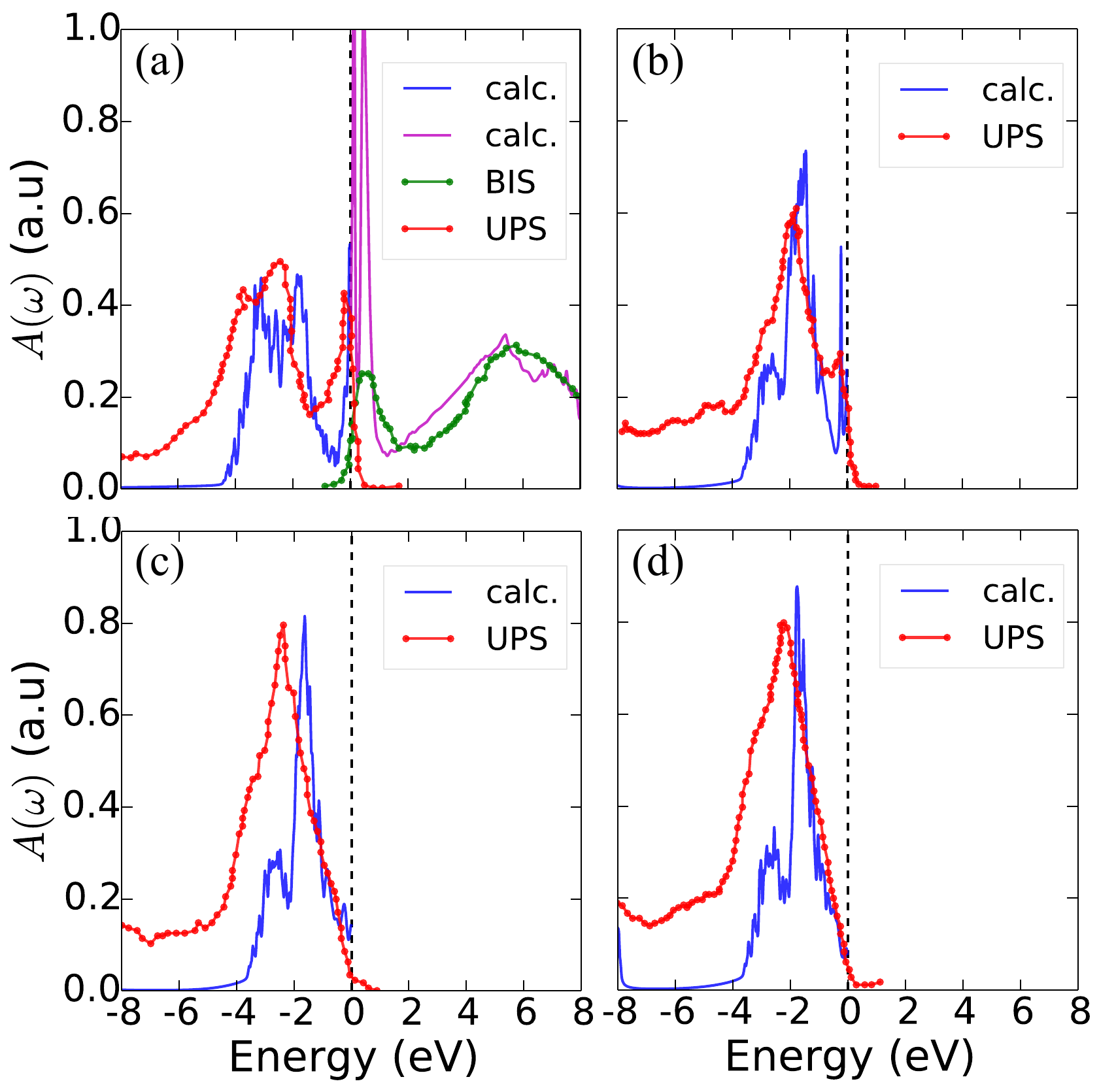}
\caption{(Color online). (a)-(d) Comparisons of theoretical and experimental density of states for CeN, CeP, CeAs, and CeSb, respectively. In panel (a), the UPS data (filled red circles) and BIS data (filled green circles) are taken from Ref.~[\onlinecite{PhysRevB.18.4433}] and Ref.~[\onlinecite{Wuilloud1985}], respectively. In panels (b)-(d), the experimental UPS data are taken from Ref.~[\onlinecite{PhysRevB.18.4433}]. The Fermi levels $E_{\text{F}}$ are represented by vertical dashed lines. Notice that the spectral data have been rescaled and normalized for a better visualization. \label{fig:tdos_exp}}
\end{figure}

\begin{figure}[th]
\centering
\includegraphics[width=\columnwidth]{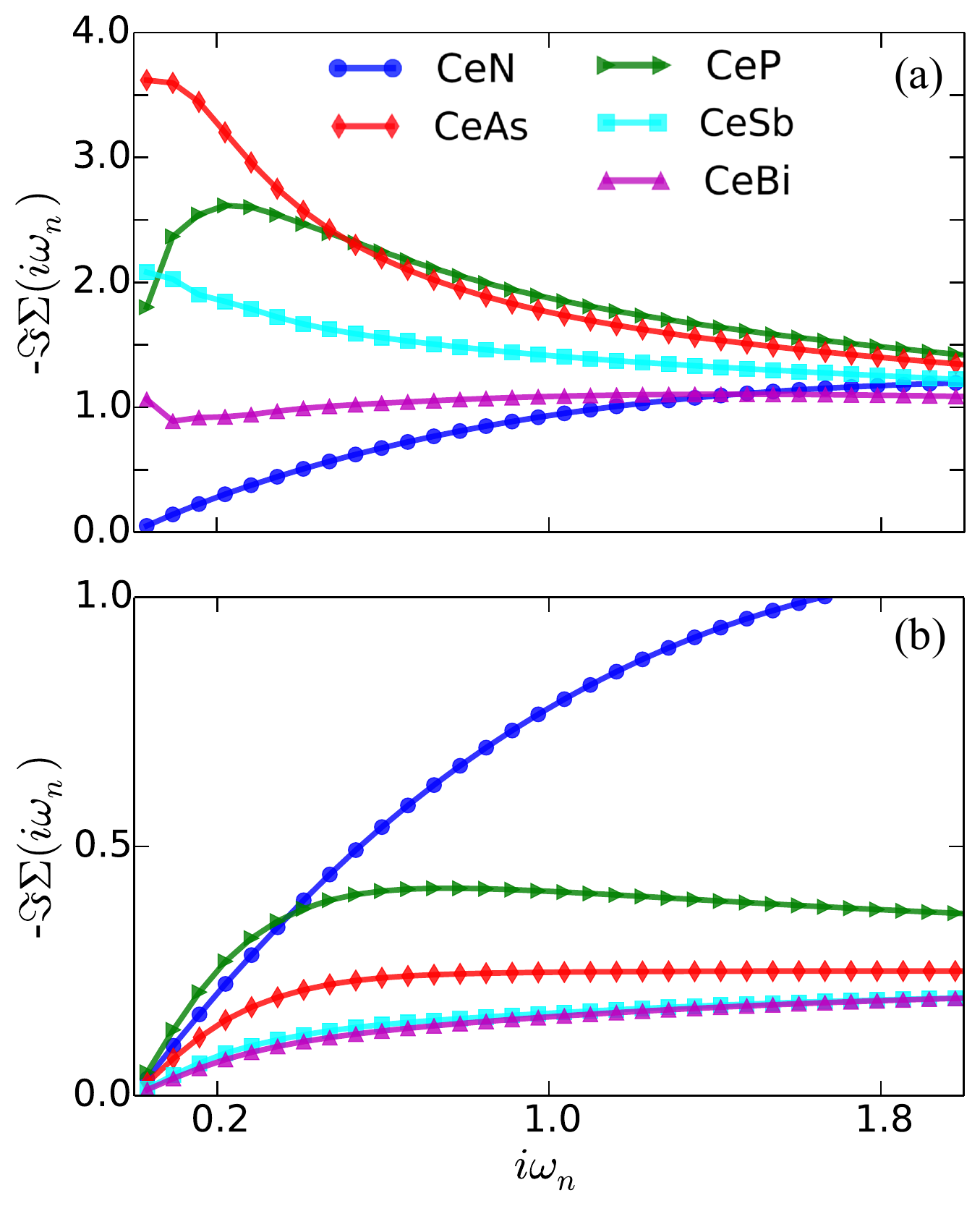}
\caption{(Color online). Imaginary parts of Matsubara $4f$ self-energy functions. (a) $4f_{5/2}$ components. (b) $4f_{7/2}$ components. The self-energy data are measured directly in the CT-HYB quantum impurity solver, instead of being calculated by Dyson's equation. \label{fig:tsigma}} 
\end{figure}

\begin{figure}[th]
\centering
\includegraphics[width=\columnwidth]{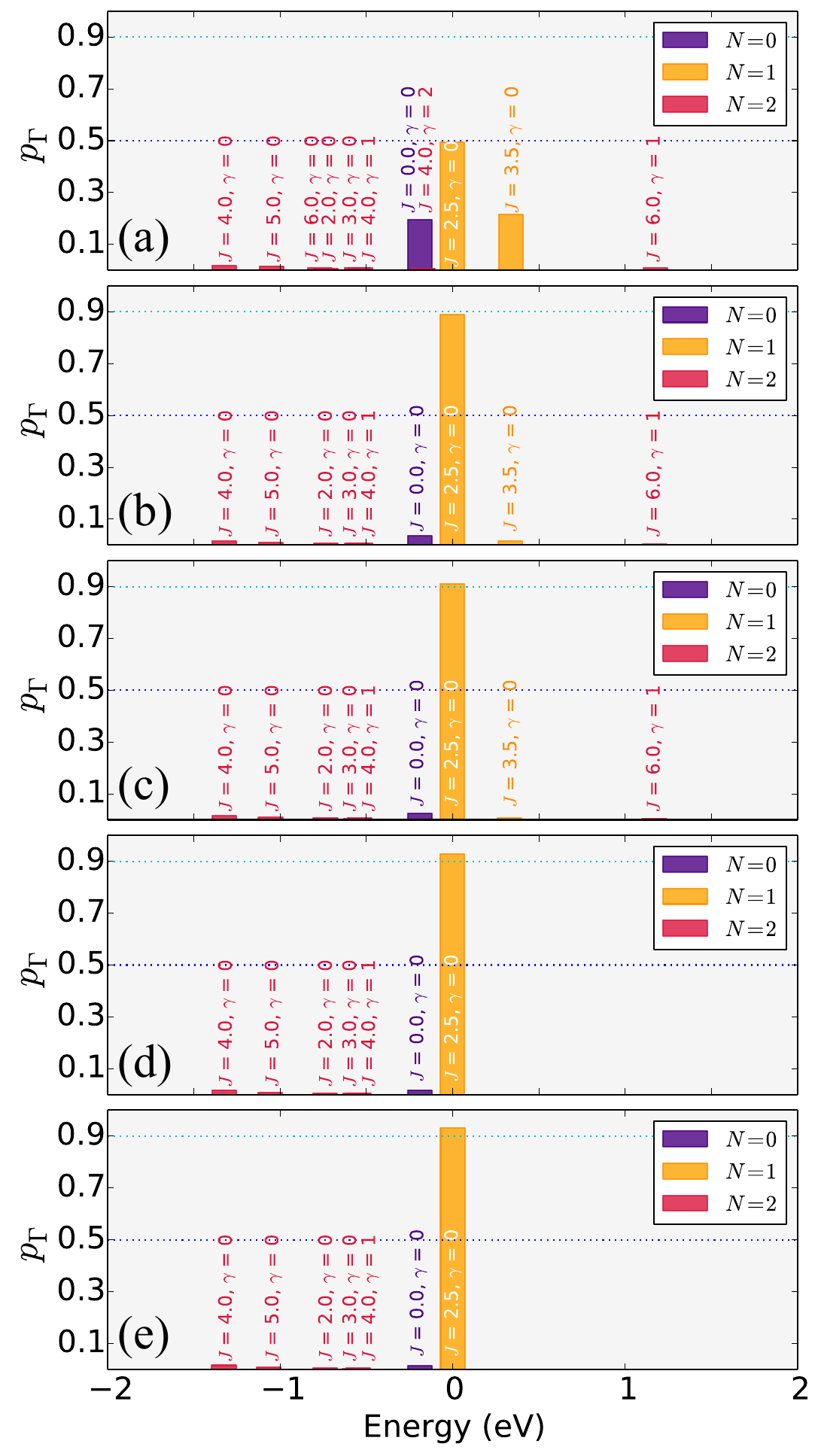}
\caption{(Color online). Atomic eigenstates histograms of Ce$X$ ($X=$ N, P, As, Sb, and Bi). (a) CeN, (b) CeP, (c) CeAs, (d) CeSb, (e) CeBi. Here we used three good quantum numbers to label the atomic eigenstates. They are $N$ (total occupancy), $J$ (total angular momentum), and $\gamma$ ($\gamma$ stands for the rest of the atomic quantum numbers, such as $J_z$). Note that the contribution from $N = 3$ atomic eigenstates is too trivial to be visualized in these panels. \label{fig:prob}}
\end{figure}

\subsection{Momentum-resolved spectral functions}

At first, we performed analytical continuation on the Matsubara self-energy functions $\Sigma(i\omega_n)$ by using the maximum entropy method. Then the obtained self-energy functions on real axis $\Sigma(\omega)$ are used to calculate the momentum-resolved spectral functions $A(\mathbf{k},\omega)$ and density of states $A(\omega)$~\cite{PhysRevB.81.195107}. 
 
The momentum-resolved spectral functions $A(\mathbf{k},\omega)$ of Ce$X$ along the high-symmetry lines $X-\Gamma-W$ in the first irreducible Brillouin zone are shown in Fig.~\ref{fig:akw}. Clearly, this series can be roughly classified into two kinds. As for CeN, the most prominent feature is the intense flat band structure near the Fermi level, which is likely from the contributions of 4$f$ orbitals. It indicates that the 4$f$ electrons in CeN are itinerant and take part in chemical bonding actively. As for CeP, CeAs, CeSb, and CeBi, the situations are somewhat different. Their spectral functions share some common characteristics: (i) For CeP and CeAs, the flat band features near the Fermi level are still discernible, but they become much dimmer and weaker than that is observed in CeN. For CeSb and CeBi, the flat bands around the Fermi level are almost invisible, implying the completely localized 4$f$ orbitals. (ii) The ligand $p$ bands are slightly renormalized and shifted toward the Fermi level as compared to those of CeN. We also notice hole pockets at the $\Gamma$-point corresponding to $X-5p$ bands, and electron pockets at the $X$-point belonging to Ce-$5d$ state~\cite{PhysRevB.56.13654}. (iii) The $4f - p$ hybridization is apparent when $\omega >$ 2.0~eV. 

\subsection{Density of states and hybridization functions}

In Fig.~\ref{fig:tdos}(a) and (b), the total and 4$f$ partial density of states of Ce$X$ are shown, respectively. For CeN, there exist sharp and strong quasiparticle resonance peaks in the vicinity of Fermi level, and a large ``hump'' between 3~eV and 8~eV. According to Fig.~\ref{fig:tdos}(b), the quasiparticle resonance peaks consist of the low-lying $4f_{5/2}$[see Fig.~\ref{fig:tdos}(b2)] and high-lying $4f_{7/2}$[see Fig.~\ref{fig:tdos}(b1)] states. The splitting energy between these two states is approximately 300~meV, which is in accordance with those measured in the other cerium-based heavy fermion compounds~\cite{PhysRevLett.108.016402,shim:1615}. The predominant contribution to the ``hump'' comes from the upper Hubbard bands. The central energy is about 4.5~eV. Since most of the 4$f$ states are unoccupied, the majority of 4$f$ spectral weights is above the Fermi level. The lower Hubbard bands are extremely weak. Concerning the rest of cerium monopnictides, the quasiparticle resonance peaks are greatly reduced. For CeSb and CeBi, these peaks nearly disappear. The upper Hubbard bands are shifted obviously to the Fermi level, which suggest again that the Ce-4$f$ orbitals become more localized and correlated when $X$ = P, As, Sb, and Bi than $X$ = N. 

Figure~\ref{fig:tdos}(c2) and (c1) depicts hybridization functions for the $4f_{5/2}$ and $4f_{7/2}$ states, respectively. It is observed that whether the $4f_{5/2}$ state or the $4f_{7/2}$ state, $-\Im \Delta(\omega = 0) / \pi$ (i.e, the 4$f$ hybridization function at the Fermi level) in CeN is always larger than those in Ce$X$ (where $X$ = P, As, Sb, and Bi). It means that when $X$ changes from N to Bi, the hybridization between Ce-$4f$ and $X$'s ligand orbitals is gradually suppressed.

The density of states of Ce$X$ has been extensively studied by using PES several decades ago. In order to verify the correctness of our calculations, we try to compare the calculated results with the available experimental data in Fig.~\ref{fig:tdos_exp}. Let's concentrate on CeN at first. The representative two-peak structure with a small shoulder peak around -2 eV is correctly reproduced by our DFT + DMFT calculations~\cite{PhysRevB.18.4433,Wuilloud1985,PhysRevB.31.6251}. For the unoccupied state, the broad ``hump'' between 3~eV and 8~eV is successfully captured. This feature is ascribed to the 4$f^{2}$ atomic multiplets. It is worth pointing out that in the previous DFT + DMFT calculations with Hubbard-I approximation as quantum impurity solver~\cite{PhysRevB.86.115116}, the authors failed to reproduce the quasiparticle resonance peaks near the Fermi level~\cite{PhysRevB.18.4433,PhysRevB.24.3651}. As for CeP, CeAs, and CeSb, the calculated results are in good agreement with the experimental spectra, including the shoulder peaks near -3~eV and the small quasiparticle resonance peak in CeP.  The small discrepancies between the theoretical and experimental spectra are likely attributed to the uncertainty in the Coulomb interaction parameters and the use of oversimplified double counting scheme~\cite{jpcm:1997}. Thus, we come to a conclusion that our DFT + DMFT calculated results are reliable and reasonable.

\subsection{Self-energy functions}

In general, the electronic correlation effect is encapsulated by the self-energy function. Traditionally, the self-energy functions can be calculated via the Dyson' equation~\cite{RevModPhys.68.13}. The resulting data are usually fluctuating and full of noise. In the present work, in order to obtain high-precision data for the self-energy functions, we try to measure them directly in the CT-HYB quantum impurity solver~\cite{PhysRevB.75.155113}. In Fig.~\ref{fig:tsigma}, the imaginary parts of 4$f$ self-energy functions are illustrated. First of all, the low-frequency parts of self-energy functions show very strong orbital differentiation. The low-frequency behaviors of the $4f_{5/2}$ and $4f_{7/2}$ states are completely different, which means that the 4$f$ electronic correlation in Ce$X$ is probably orbital dependent. It is not at all surprised because this phenomenon has been identified in many cerium-based heavy fermion materials~\cite{PhysRevB.98.195102,PhysRevB.94.075132} and strongly correlated $5f$ electron systems~\cite{PhysRevB.99.045109} a few years ago. Second, the 4$f$ self-energy functions of CeN are quite distinctive from those of the other cerium monopnictides. For example, the low-frequency part of $4f_{5/2}$ state of CeN exhibits remarkable quasi-linear behavior. It signifies a (heavy) Fermi-liquid state. However, the corresponding parts of Ce$X$ are convex ($X =$ P and As) or concave ($X = $Sb and Bi). Third, the intercept of self-energy function in $y$-axis is approximately zero for CeN. While for the other cerium monopnictides, the intercepts are finite. It means that the low-energy scattering of 4$f$ electrons in CeN is much smaller than those in the rest of cerium monopnictides. Fourth, the low-energy scattering of the $4f_{7/2}$ states is usually smaller than that of the $4f_{5/2}$ states.  

\begin{table}[th]
\caption{The effective electron mass $m^\star$ and quasi-particle weight $Z$ of the $4f_{5/2}$ and $4f_{7/2}$ states for Ce$X$ (where $X$ = N, P, As, Sb, and Bi). Here, $m_e$ means the bare electron mass. \label{tab:sig}}
\begin{ruledtabular}
\begin{tabular}{ccccc}
      & \multicolumn{2}{c}{$4f_{5/2}$} & \multicolumn{2}{c}{$4f_{7/2}$} \\
cases & $m^{\star}/m_e$ & $Z$ & $m^{\star}/m_e$ & $Z$ \\
\hline
CeN     & 02.603 & 0.384  & 2.088 & 0.479  \\
CeP     & 58.331 & 0.017  & 2.502 & 0.400  \\
CeAs    & 116.185 & 0.009  & 1.829 & 0.547  \\
CeSb    & 67.248 & 0.015  & 1.462 & 0.684  \\
CeBi    & 34.830 & 0.029  & 1.385 & 0.722  \\
\end{tabular}
\end{ruledtabular}
\end{table}

Based on the self-energy data, the quasi-particle weight $Z$ and effective electron mass $m^{\star}$ can be evaluated via the following equation~\cite{RevModPhys.68.13}: 
\begin{equation}
Z^{-1} = \frac{m^{\star}}{ m_e} \approx 1 - \frac{\Im\Sigma(i\omega_0)}{ \omega_0},
\end{equation}
where $\omega_0 = \pi / \beta$ and $m_e$ denotes the mass of non-interacting band electron. 
The calculated $Z$ and $m^{\star}$ are summarized in Table~\ref{tab:sig}. We find that the $4f_{7/2}$ states are less renormalized. Its $Z \approx$ 0.4-0.7 and $m^{\star} \approx $ 1.3-2.5 $m_{e}$. However, the $4f_{5/2}$ states are strongly renormalized. Notice that CeAs exhibits the largest $m^{\star}$ and smallest $Z$, implying that its 4$f$ electrons are probably the most localized. Since the ratio $R \equiv Z(4f_{7/2}) / Z(4f_{5/2})$ is so large ($R \sim$ 1.25 for CeN, and $R > 20$ for CeP, CeAs, CeSb, and CeBi), it is concluded that these materials are in the so-called orbital-selective heavy fermion state, or equivalently, orbital-selective localized state~\cite{PhysRevB.98.195102}. 

\subsection{Valence state fluctuations}

Valence state fluctuation or mixed-valence behavior is a common feature in many cerium-based heavy fermion materials~\cite{PhysRevB.98.195102}. In the present work, by utilizing the atomic eigenvalue probability $p_{\Gamma}$, which stands for the probability to find out a 4$f$ valence electron in a given atomic eigenstate $|\psi_\Gamma\rangle$, we can make a reliable estimation about the magnitude of valence state fluctuation in Ce$X$. The CT-HYB quantum impurity solver is capable of recording the atomic eigenvalue probability $p_{\Gamma}$~\cite{PhysRevB.75.155113}. In Fig.~\ref{fig:prob}, the calculated results for Ce$X$ are illustrated as histograms. Here, the atomic eigenstates $|\psi_\Gamma\rangle$ are labelled by using some good quantum numbers such as total occupation $N$ and total angular momentum $J$.

We discover that the valence state histograms of CeN [see Fig.~\ref{fig:prob}(a)] are quite different from those of the other four compounds. The probability for the atomic eigenstate $|N = 1, J = 2.5, \gamma = 0\rangle$ is about 50\%. Evidently, it is the predominant atomic eigenstate. The following atomic eigenstates are  $|N = 1, J = 3.5, \gamma = 0\rangle$ and $|N = 0, J = 0.0, \gamma = 0\rangle$. Their probabilities account for 22\% and 20\%, respectively. The probabilities for the rest atomic eigenstates are negligible. Therefore, it is suggested that the $4f$ electrons in CeN favor to fluctuate among the above three principle competing atomic eigenstates and become itinerant through hybridization with ligand electrons. When $X$ changes from N to P, As, Sb, and Bi, the corresponding atomic eigenstate probability for $|N = 1, J = 2.5, \gamma = 0\rangle$ soars from 50\% to 90\%. At the same time, the atomic eigenstates probabilities for $|N = 0, J = 0.0, \gamma = 0\rangle$ and $|N = 1, J = 3.5, \gamma = 0\rangle$ decrease rapidly. For CeP and CeAs, they account for less than $ 4\%$. For CeSb and CeBi, they are less than 1\% and are nearly invisible in Fig.~\ref{fig:prob}(d) and (e). It seems that the 4$f$ electrons in CeP, CeAs, CeSb, and CeBi are very localized, and virtually confined to the primary atomic eigenstate $|N = 1, J = 2.5, \gamma = 0\rangle$. Meanwhile, the corresponding valence state fluctuations are very weak. In short, the redistribution of atomic eigenstates probabilities strongly relies on the atomic number of $X$.

By summing up the atomic eigenstates probabilities $p_{\Gamma}$ with respect to $N$, we can derive the distribution of 4$f$ electronic configurations. It will provide further information about the 4$f$ valence state fluctuations and mixed-valence behaviors. In CeN, on one hand, the $4f^{1}$ configuration is predominant and its probability is about 70\%. On the other hand, the probabilities of the $4f^{0}$ and $4f^{2}$ configurations are about 20\% and 9.2\%, respectively. It indicates the mixed-valence nature of CeN, which accords with the findings of previous experiments~\cite{PhysRevLett.39.956,PhysRevB.18.4433,PhysRevB.24.3651,PhysRevB.42.8864,WACHTER2013235,Wuilloud1985}. For $X$ = P, As, Sb, and Bi, the $4f^{1}$ configuration actually becomes more overwhelming. Its probability is larger than 90\%, while those of the $4f^{2}$ and $4f^{0}$ configurations decline to less than $ 6\%$. It means that the $4f$ valence state fluctuation in CeN is the most remarkable. When $X$ grows from N to Bi, the 4$f$ valence state fluctuations will be greatly suppressed. In consequence, the mixed-valence behaviors will become very trivial.


\section{Discussions\label{sec:dis}}

In this section, we would like to discuss some important issues and questions.

\emph{4f itinerant-localized crossover or transition.} According to the momentum-resolved spectral functions and density of states, we believe that the 4$f$ electrons in CeN is itinerant, while they tend to be localized in CeP, CeAs, CeSb, and CeBi. In other words, the 4$f$ itinerant-localized crossover may occur between CeN and CeP. Provided that the N atoms in CeN are substituted gradually by $X$ atoms ($X$ = P, As, Sb, or Bi), a $4f$ itinerant-localized crossover is naturally expected. Then a new question rises: what's the driving force of this crossover? First, when $X$ grows from N to Bi, the lattice constants of Ce$X$ increase monotonously~\cite{Duan2007Electronic}. The unit cell volume of CeBi is almost twice of the one of CeN (see Table~\ref{tab:param}). The larger Ce-Ce bond length is, the more localized the 4$f$ electrons become. Second, we think that the spin-orbital coupling effect of $X$'s $p$ orbitals should play a nontrivial role in this crossover. Generally, the spin-orbital coupling is stronger for the heavier elements, where the electrons acquire large velocities near the nucleus. So, $\lambda_{\text{Bi,6p}} > \lambda_{\text{Sb,5p}} > \lambda_{\text{As,4p}} > \lambda_{\text{P,3p}} > \lambda_{\text{N,2p}}$, where $\lambda$ denotes the strength of spin-orbital coupling. Thus, the hybridization between Ce's 4$f$ and $X$'s $np$ ($n = 2\sim6$) orbitals should be tuned inevitably by the spin-orbital coupling. Notice that this mechanism is quite similar to the TKI-WKSM transition observed in Ce$_{3}$Bi$_{4}$(Pt$_{1-x}$Pd$_{x}$)$_{3}$ series, which is actually driven by the difference in spin-orbital coupling between Pt and Pd atoms~\cite{Lai93,PhysRevLett.118.246601,zhu2019}. 

\emph{Evolution of 4$f$ electronic structures in Ce$X$.} The 4$f$ electronic structures of Ce$X$ share some common features. The 4$f$ electrons are all correlated. The electronic correlations are orbital dependent, i.e, the $4f_{5/2}$ states are much more correlated than the $4f_{7/2}$ states. The evidence is that the quasiparticle weight $Z$ (or effective electron mass $m^{\star}$) of the $4f_{5/2}$ states is much smaller (or larger) than the one of the $4f_{7/2}$ states. On the other hand, the 4$f$ electronic structures of CeN differ from all the other Ce$X$ compounds obviously. The 4$f$ electrons in CeN are itinerant, with strong valence state fluctuation. We can observe the quasiparticle resonance peak in the Fermi level and Fermi-liquid-like behavior in the low-frequency parts of 4$f$ self-energy functions. Conversely, the 4$f$ electrons in the other Ce$X$ compounds are totally localized. The $4f-p$ hybridization near the Fermi level and valence state fluctuation are rather weak. Apparently, their self-energy functions deviate from the description of Landau Fermi-liquid theory. They are also not mixed-valence compounds under ambient condition, though there are some experimental evidences that these compounds might come to be mixed-valence under moderate pressure~\cite{PhysRevLett.36.366}.    

\emph{Electronic correlation and magnetism in Ce$X$.} Usually the ground states of Ce$X$ ($X =$ P, As, Sb, and Bi) are antiferromagnetic, except for the paramagnetic CeN~\cite{Duan2007Electronic}. It is easy to understand because the 4$f$ electrons in Ce$X$ are localized and tend to form local moments. Among these compounds, CeSb and CeBi are well known due to their complicated magnetic phase diagram under pressure or under magnetic field~\cite{Kohgi2000Physics,Hulliger1975Low,Fischer1978Magnetic,Meier1978Magnetic}. Previous studies suggested that these unusual magnetic properties originate from the small crystal field splitting. Due to the 4$f$ localization and $4f-p$ mixing effect, the crystal field excited state with $\Gamma_8$ character is pulled down below the crystal field ground state $\Gamma_7$. Then stacking magnetic structures are formed with strongly polarized $\Gamma_8$ Ce layer and paramagnetic $\Gamma_7$ Ce layer. This scenario looks good, but it requires the cubic symmetry. However, when CeSb and CeBi transform from paramagnetic phase to ordered phases, their crystal structures also distort from the cubic ones to the tetragonal ones~\cite{Duan2007Electronic}. Furthermore, their lattice constants in the tetragonal structures (along $a$-axis and $c$-axis) diminish with decreasing temperature when $T < T_{\text{N}}$. Thus, the above model may be not enough to explain the atypical magnetic properties of Ce$X$. A credible model for this problem should at least take the temperature-dependent crystal structures and the corresponding crystal field splitting into considerations. Anyway, we anticipate that the 4$f$ electronic correlations should play a vital role in the electronic structures and magnetic properties of CeSb and CeBi with tetragonal symmetry. More DFT + DMFT calculations are being undertaken. 


\section{conclusion\label{sec:summary}}

In summary, the $4f$ electronic structures of cerium-based monopnictides Ce$X$ ($X$=N, P, As, Sb, and Bi) have been systematically investigated by using the DFT + DMFT approach. The momentum-resolved spectral functions $A(\mathbf{k},\omega)$, total and $4f$ partial density of states $A(\omega)$ and $A_{4f}(\omega)$, hybridization functions, Matsubara self-energy functions, and $4f$ valence state fluctuations are studied. The calculated results are consistent with the available experimental data. However, since the experimental data are very limited, most of the calculated results act as useful predictions. 

It is confirmed that $4f$ states of CeN are the most itinerant among the five compounds and display mixed-valence behavior. When $X$ goes from N to Bi, the $4f$ electrons in Ce$X$ turn out to be more and more localized. It is proposed that the 4$f$ itinerant-localized crossover probably takes place between CeN and CeP, which is accompanied by vanishing of quasiparticle resonance peak in the Fermi level and regression of $4f$ valence state fluctuation. In particular, the orbital-dependent $4f$ correlations are identified in Ce$X$. Their $4f_{5/2}$ orbitals are more correlated and more renormalized than the $4f_{7/2}$ orbitals, which is in analogy with the other cerium-based heavy fermion compounds. Finally, we would like to point out that the 4$f$ electronic structure is tightly connected with the magnetism of Ce$X$. In order to interpret the intricate magnetic orderings in CeSb and CeBi, which remains a long standing issue and yet to be answered, a deep understanding about the 4$f$ electronic structures of Ce$X$ is indispensable. The present study about Ce$X$'s 4$f$ electronic structures fills in this gap, and enriches our knowledge about the exotic properties of cerium-based heavy fermion compounds. However, further experimental and theoretical validations are still highly desired.

\begin{acknowledgments}
This work was supported by the Science Challenge Project of China (No.~TZ2016004), and the Natural Science Foundation of China (No.~11704347 and No.~11874329). The DFT + DMFT calculations were performed on the Sugon cluster (in the Institute of Physics, CAS, China).
\end{acknowledgments}


\bibliography{CeX}

\end{document}